\documentclass[11pt,twoside]{article}
\usepackage{asp2010}
\usepackage{natbib}

\resetcounters

\aspvolume{471} 
\aspvoltitle{Origins of the expanding universe: 1912-1932}
\aspcpryear{2013} 
\aspvolauthor{J.A. Peacock}

\def\kms{{\,\rm km\,s^{-1}}}
\def\kmsmpc{{\,\rm km\,s^{-1}Mpc^{-1}}}
\def\mpcoh{{\,h^{-1}\,\rm Mpc}}

\def\m@th{\mathsurround=0pt }
\def\eqalign#1{\null\,\vcenter{\openup1\jot \m@th
 \ialign{\strut\hfil$\displaystyle{##}$&$\displaystyle{{}##}$\hfil
 \crcr#1\crcr}}\,}

\def\be{\begin{equation}}
\def\ee{\end{equation}}
\def\citejap#1{\citeauthor{#1}\ \citeyear{#1}}

\begin{document}

\title{Slipher, galaxies, and cosmological velocity fields}
\author{John A. Peacock
\affil{Institute for Astronomy, University of Edinburgh, Royal Observatory, Edinburgh EH9 3HJ, UK}
}

\begin{abstract}
By 1917, V.M. Slipher had singlehandedly established a general
tendency for `spiral nebulae' to be redshifted (21 out of 25
cases). From a modern perspective, it could seem surprising that the
discovery of the expansion of the universe was not announced at this
point. Examination of the data and arguments contained in Slipher's
papers shows that he reached a more subtle conclusion: the
identification of cosmological peculiar velocities, including the bulk
motion of the Milky Way, leading to a beautiful argument in favour of
spiral nebulae as distant stellar systems.  Nevertheless, Slipher's
data actually contain evidence at $>$$8\sigma$ for a positive mean
velocity, even after subtracting the best-fitting dipole pattern owing
to motion of the observer. In 1929, Hubble provided distance estimates
for a sample of no greater depth, using redshifts due almost entirely
to Slipher. Hubble's distances turned out to be flawed in two
distinct ways: in addition to an incorrect absolute calibration, the
largest distances were systematically under-estimated. Nevertheless,
he claimed the detection of a linear distance-redshift relation.
Statistically, the evidence for such a correlation is less strong than
the simple evidence for a positive mean velocity in Hubble's sample.
Comparison with modern data shows that a sample of more than twice Hubble's
depth would generally be required in order to reveal clearly the
global linear expansion in the face of the `noise' from peculiar velocities.
When the theoretical context of the time is examined, the role of the
de Sitter model and its prediction of a linear distance-redshift
relation looms large.  A number of searches for this relation were
performed prior to Hubble over the period 1924--1928, with a similar
degree of success. All were based on the velocities measured by
Slipher, whose work from a Century ago stands out both for the
precision of his measurements and for the subtle clarity of the
arguments he employed to draw correct conclusions from them.
\end{abstract}

\section{Introduction}

This talk does not pretend to be a professional exercise in the history
of science, but all cosmologists have a fascination with origins:
thus we should expect to have a reasonable familiarity with the developments
that set our subject in motion. There is indeed a conventional narrative that
has been repeated in compressed form in innumerable classrooms and
public talks -- generally centring on \citet{Hubble1929} as having provided the first observational
evidence for an expanding universe. But this conference
represents the convergence of many individual trajectories
of re-evaluation, all reflecting a growing recognition that our standard tale
is seriously at variance with the actual events.

Chief among the casualties of this over-simplification has been V.M. Slipher.
His existence was not hidden, and he appears on the first page
of the textbook by \citet{Peebles1971}, which was a great influence on my
generation of cosmologists. But a proper appreciation of Slipher's work was 
perhaps hindered by the fact that
all of his major papers appeared in the obscurity
of internal Lowell reports, or journals that ceased publication.
In 2004, I tried and failed to discover electronic versions of
Slipher's seminal papers anywhere on the web. But the Royal Observatory Edinburgh
is fortunate enough to possess an outstanding collection of
historical journals, so I was able to track down the originals and make scans.
Since then, I am proud to say that my
complaints to ADS have had an effect: not only are Slipher's main
papers all now listed (Slipher's 1917 masterpiece lacked any entry whatsoever in the
database), but you can also find the scans I made through ADS.

Reading these papers cannot fail to generate an enormous respect
for Slipher as a scientist:
they are confidently argued, and make some
points that are astonishingly perceptive with the aid of 21st-century
hindsight. Given the confusion that naturally attended the first engagement
with modern cosmological questions, this is all the more impressive. When
all is unclear, it is tempting to hedge papers with so many qualifications
that no conclusion ever emerges -- but the mark of a great scientist is to
stick your neck out and state firmly what you believe to be true.
Slipher achieves this on a number of occasions, and his conclusions
have stood the test of time.

The intention of this presentation is to try to illuminate
the magnitude of Slipher's achievements by viewing them through the eyes of a working
cosmologist, and going back to the analysis of the original data.
In particular, by comparing with modern data, the aim is to understand
why Slipher did not use the general tendency for galaxies to be
redshifted as evidence for the expansion of the universe -- but how
he came to reach an under-appreciated conclusion of similar importance.

\section{Slipher's great papers}

Before focusing on Slipher's most important paper (\citejap{Slipher1917}), it
is worth giving a brief overview of his achievements during the
period when he was the lone pioneer of `nebular' spectroscopy.
During all of this, it should be clearly borne in mind that the
nature of the nebulae was unclear in this period; although the
`island universe' hypothesis of distant stellar systems was
a known possibility, a considerable weight of opinion viewed
the spiral nebulae as planetary systems in formation.

In \citet{Slipher1913} the blueshift of Andromeda was measured to be $300\kms$.
This velocity was very high by the standards of the time, and there
could understandably be skepticism about whether this really
was a Doppler shift (cf. the quasar
redshift controversy). Slipher trenchantly asserts that ``\dots we have
at the present no other interpretation for it. Hence we may conclude
that the Andromeda Nebula is approaching the solar system\dots''. Since
the blue shift is now believed to be induced by dark-matter density
perturbations, it is amusing to note Slipher's speculation that the
nebula ``might have encountered a dark star''.

\citet{Slipher1914} was unknown to me prior to my 2004 archival search, but
appears to be the first demonstration that spiral galaxies rotate.
This would make Slipher a figure of importance, even if he had done
nothing else. A striking contrast with modern `publish or perish'
culture is Slipher's statement that he believed he had data showing
the tilt of spectral lines, but was not fully satisfied; therefore he
waited an entire year until he could repeat and check the results.

The paper in which Slipher
presented his results to the American Astronomical Society 
(\citejap{Slipher1915}; August
1914 meeting) is perhaps the most well-known.
Out of 15 galaxies, 11 were clearly redshifted, and he
received a standing ovation after reporting this fact. 
It must have been clear to all present
that this was an observation of deep significance -- even if the
interpretation was lacking at the time.

The 1917 paper is the most extensive of Slipher's works on
nebular spectroscopy, but surprisingly it seems to
be less well known than the papers of 1913 and 1915, and
I had never seen any mention of its contents prior to reading it
for the first time in 2004. The redshift:blueshift
ratio has now risen to 21:4, but it is the interpretation that is
startling.

\section{Slipher's intellectual leap of 1917}

Although the mean redshift of the 1917 sample is large and positive, Slipher does
not draw what might today be regarded as the obvious conclusion:

\medskip
{\narrower\noindent
The mean of the velocities with regard to sign
is positive, implying that the nebulae are receding
with a velocity of nearly 500 km.
This might suggest that the spiral nebulae are
scattering but their distribution on the sky is not
in accord with this since they are inclined to cluster.

}

\medskip
The term ``scattering'' clearly denotes a tendency to recede in
all directions, which must be regarded as the most basic symptom
of an expanding universe. The reason Slipher does not
state this as a conclusion is because there is an issue of
reference frame. Astronomers of this era were completely familiar
with the fact that the Sun moves with respect to the nearby
stars, inducing a dipole pattern in the observed velocities. It
must therefore have seemed entirely natural to fit a dipole pattern
to the sky distribution of velocities.
Slipher makes this analysis, deducing a mean velocity of $700\kms$
for the Sun and thus
noting that we are not at rest with respect to
the other galaxies on average. He then
makes a tremendous intellectual leap, which is described in language
of a beautiful clarity:

\begin{figure}[ht]
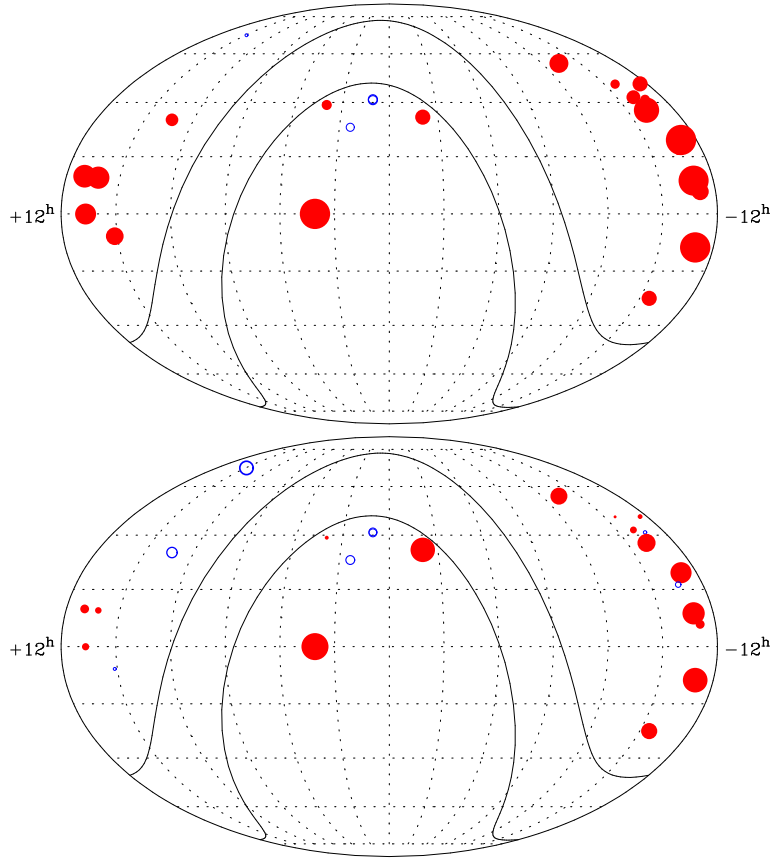

\centering
\begin{tabular}{c}
\includegraphics[scale=0.7]{moll.eps}\\
\includegraphics[scale=0.7]{moll2.eps} \\
\end{tabular}
\caption{The sky distribution of Slipher's 1917 galaxies, in Mollweide
projection of celestial coordinates, indicating the galactic plane at
$b=\pm15^\circ$.  Velocities covering the range $-300\kms$ to $+1000\kms$ 
are coded red solid (redshift) or blue open (blueshift), with the width of the symbol
being proportional to the magnitude of the velocity. The top panel shows the
observed velocities; the lower panel shows the same data after removal of the
best-fitting dipole.  Removal of the dipole reduces the mean velocity from
$502\kms$ to $145\kms$ with a dispersion of $414\kms$.}\label{peacockfig01}
\end{figure}

\medskip
{\narrower\noindent
 We may in like manner determine our motion relative to the spiral
  nebulae, when sufficient material becomes available. A preliminary
  solution of the material at present available indicates that we are
  moving in the direction of right ascension 22 hours and declination
  $-22^\circ$ with a velocity of about 700 km.  While the number of
  nebulae is small and their distribution poor this result may still
  be considered as indicating that we have some such drift through
  space.  For us to have such motion and the stars not show it means
  that our whole stellar system moves and carries us with it. It has
  for a long time been suggested that the spiral nebulae are stellar
  systems seen at great distances \dots This theory, it seems to me,
  gains favor in the present observations.

}

\medskip
This argument is a dizzying shift of perspective: we start in the
Milky Way looking out at the nebulae, from whose dipole reflex motion
Slipher correctly infers that the entire Milky Way is in motion at a
previously undreamed-of speed. This is almost as shocking a discovery
as Copernicus's proposal that the Earth is in motion. But then the
perspective shifts, and suddenly Slipher imagines himself to be within
one of the nebulae -- looking out at the Milky Way and other nebulae:
since they all have rms motions in the region of $400\kms$, they must
clearly all be the same kind of thing.  Hence the nebulae are hugely
distant analogues of the Milky Way.  This, remember, is 8 years before
Hubble detected Cepheids in Andromeda and settled the `island
universe' question directly. Slipher was not actually the first to
consider measuring the motion of the Milky Way in this fashion (see
\citejap{Sullivan1916}; \citejap{Truman1916}; \citejap{Young1916}; 
\citejap{Paddock1916}; \citeauthor{Wirtz1916} \citeyear{Wirtz1916},
\citeyear{Wirtz1917}; I thank Michael Way for pointing out these
references). But all these investigations used Slipher's data, and
somehow their conclusions lack his confidence and compact clarity
regarding the physical implications -- although it would be 
interesting to know if he was motivated by these earlier papers.

Given the neatness of the argument that Slipher uses here, one can
hardly complain that he does not focus on the fact that the mean
redshift is non-zero, even after adjustment for the best-fitting
dipole. Indeed, this feature is not statistically compelling: the mean
redshift after dipole subtraction is $145\kms$ with an rms of
$414\kms$, which is only a $1.8\sigma$ deviation from zero. This transformation
of the data can
be seen at work in the sky distributions of Slipher's data shown in
Figure \ref{peacockfig01}: 
the largest velocities are concentrated around $\alpha=12^h$,
$\delta=0^\circ$ (dominated by the Virgo cluster), and so can be
heavily reduced by an appropriate dipole -- even though the pattern of
residuals shown in the lower panel of Figure \ref{peacockfig01}
is clearly non-random.

It is a great pity that Slipher
did not revisit this analysis with the redshifts he continued
to accumulate. Rather than write a further paper, he was content
simply to have these results appear in Eddington's (\citeyear{Eddington1923})
book (page 162). By this time, there were 41 velocities, of which 36 were
positive; the most negative remained at the $-300\kms$ of M31,
whereas 5 objects had redshifts above $1000\kms$, including
$1800\kms$ for NGC584, which came close to doubling the maximum
velocity of the 1917 data. If we repeat Slipher's 1917 analysis
with the expanded 1923 dataset, the mean velocity after subtracting
the best-fitting dipole rises to $201\kms$ with an
rms of $508\kms$; this is now a $2.5\sigma$ deviation from zero, 
and so Slipher's velocities alone give a very clear signal of
a general tendency towards expansion (in fact, for reasons explained
below, this analysis greatly underestimates the significance of the effect).
Eddington does not attempt a dipole analysis, but his discussion
of Slipher's data clearly focuses on the high mean value of the raw data as
representing a general tendency for galaxies to be redshifted, albeit
with some dispersion. However, this is not simply an abstract statement about
the pattern of the numbers: for reasons explained in the
following section,  Eddington had a theoretical
expectation of a general redshift.

\section{The theoretical prior}

By the time of Slipher's 1917 analysis, the theorists 
were on the march. Two years after the
creation of General Relativity, \citet{Einstein1917} had created his
static cosmological model, introducing the cosmological
constant for the purpose. This is a wonderful paper, which can be
read in English in e.g. \citet{Bernstein1986}, and the basic
argument is one that Newton might almost have generated.
Consider an infinite uniform sea of matter, which we want to
be static (an interesting question is whether Einstein was
influenced by data in imposing this criterion, or whether
he took it to be self-evident): we want zero gravitational
force, so both the gravitational potential, $\Phi$ and
the density, $\rho$ have to be constant. The trouble is,
this is inconsistent with Poisson's equation, $\nabla^2\Phi
= 4\pi G \rho$. The `obvious' solution (argues Einstein) is
that the equation must be wrong, and he proposes instead
\be
\nabla^2\Phi + \lambda\Phi = 4\pi G \rho,
\ee
where $\lambda$ has the same logical role as the 
$\Lambda$ term he then introduces into the field equations.
In fact, this is not the correct static Newtonian limit of
the field equations, which is
$\nabla^2\Phi + \Lambda = 4\pi G \rho$. But either
equation solves the question posed to Newton by
Richard Bentley concerning the fate of an infinite
mass distribution; Newton opted for a static model despite
the inconsistency analysed above:

\medskip
{\narrower\noindent \dots it seems to me that if the matter of our sun
  and planets, and all the matter of the universe, were evenly
  scattered throughout all the heavens, and every particle had an
  innate gravity towards all the rest, and the whole space, throughout
  which this matter was scattered, was but finite; the matter on the
  outside of this space would by its gravity tend towards all the
  matter on the inside, and by consequence fall down into the middle
  of the whole space, and there compose one great spherical mass. But
  if the matter was evenly dispersed throughout an infinite space, it
  would never convene into one mass\dots

}

\noindent
(see e.g. pp. 94-105 of \citejap{Janiak2004}).
With the advantage of hindsight, Newton seems tantalisingly close at this time
(10 December 1692) to anticipating Friedmann by over
200 years and predicting a dynamical universe.

But at almost the same time as Einstein's work,
the first non-static cosmological
model was enunciated by \citet{deSitter1917} -- based on the same $\Lambda$ term that
was intended to ensure a static universe. It is interesting
to compare the original forms of the metric in these two
models, as they are rather similar:
\begin{eqnarray}
{\rm Einstein:}\quad d\tau^2 &=& - dr^2 - R^2\sin^2(r/R)d\psi^2 + dt^2 \\
{\rm de Sitter:}\quad d\tau^2 &=& - dr^2 - R^2\sin^2(r/R)d\psi^2 + \cos^2(r/R)dt^2
\end{eqnarray}
Staring at these naively, it it tempting to conclude that clocks
slow down at large distances, $\propto \cos(r/R)$, where $R$ is
a characteristic curvature radius of spacetime. In this
case, a redshift-distance relation would be predicted to be
$z\simeq r^2/2R^2$ i.e. quadratic in distance. But we have 
made an unjustified assumption here, which is that a free
particle (or galaxy) will remain at constant $r$, which we
know does not actually happen. For this reason, the 
correct redshift-distance relation is linear at lowest order. This was
first demonstrated by \citeauthor{Weyl1923} (\citeyear{Weyl1923}; the 5th
edition of his book -- frustratingly, the common Dover reprint is the
4th edition). This was also shown independently by
\citet{Silberstein1924} and \citeauthor{Lemaitre1927}
(\citeyear{Lemaitre1927}; see \citejap{Lemaitre1931} for an English
translation). Interestingly, Eddington (\citeyear{Eddington1923}) proves (on
page 161) that test particles near the
origin experience an outward acceleration proportional to
distance, and from his discussion he clearly sees that this
motion will make a contribution to the observed redshift --
but he never clearly states that the leading effect is thus
a linear term in $D(z)$.

News of this prediction seems to have spread
rapidly, and there were soon a number of attempts
to look for a linear relation between redshift and distance. 
It should be made clear that no-one at this stage was thinking
about an expanding universe (Friedmann was perhaps an
exception, but he was decoupled from the interplay between
theory and experiment in the West). The aim was to
search for the `de Sitter effect' and thus
`measure the radius of curvature of spacetime'.
This game can be played with any set of objects where
radial velocities exist, together with some indicator of
distance.

\begin{figure}[ht]
\center{\includegraphics[scale=0.5]{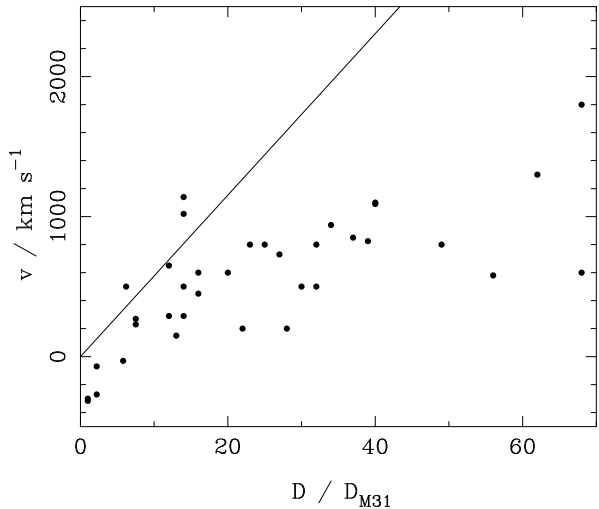}}
\caption{\citet{Lundmark1924} searched systematically for a linear
distance-redshift relation, using a variety of classes of
astronomical object. His most impressive result was obtained
using nebulae. Lacking any direct distance estimates for these,
he took a standard-candle approach, in which relative distances
were measured using the apparent magnitudes and/or diameters of
nebulae; results were quoted in units where the distance to
M31 was unity. The solid line shows the modern truth, assuming
$H_0=73\kmsmpc$ and a distance to M31 of 0.79 Mpc. It can be seen
that Lundmark's approach works reasonably well out to
$D/D_{\rm M31}\simeq 25$, but thereafter comes adrift as dwarf
galaxies are assigned incorrectly high distances (8 further dwarfs
exist at larger values of $D$, and these are not shown).}\label{peacockfig02}
\end{figure}

A number
of people (\citejap{Silberstein1924};  \citejap{Wirtz1924}; 
\citejap{Lundmark1924}) tried this, and the paper by
Lundmark is particularly impressive and comprehensive. 
In general, distances to galaxies were lacking
at that time, although the detection of Novae in M31 had suggested a
distance around 500 kpc, which is not too far off. What Lundmark did
was to assume that galaxies were standard objects; thus he was able to estimate distances
in units of the M31 distance, based on diameters and on apparent
magnitudes (these agree reasonably well). The distances clearly
correlate with Slipher's redshifts, as shown in Lundmark's Figure
5 (recreated here as Figure \ref{peacockfig02}). Lundmark was not as impressed
with his result as perhaps he ought to have been: 
``\dots we find that there may be a relation between the
two quantities, although not a very definite one''.
But despite the scatter, a positive correlation
of distance and redshift does exist, of a significance so
obvious that it hardly needs formal quantification.
Thus by 1924 it was clear that radial velocities tended to be positive,
and to increase with distance, even if it was not possible
to say with any confidence that the redshifts scaled linearly with distance.

In any case, we reiterate that the physical understanding of the
meaning of any distance-redshift relation still
had some way to go in 1924. Despite Eddington's insight
that there was a kinematical effect at work, the common
interpretation of the de Sitter model in the above papers 
was the static view that redshifts simply probed the
curvature of spacetime. And even in 1929 Hubble would mention
``the de Sitter effect'' and Eddington's argument for a kinematical
contribution without actually saying that expansion
dominates locally.

\section{Comparison with modern data}

How well might the studies of a Century ago be expected to work with
modern data? Today, we can measure relative distances to a
precision of order 5\% using Cepheid variable stars out to $D\simeq 30\mpcoh$,
or out to almost arbitrary distances using SNe Ia (with the aid of the Hubble 
Space Telescope in both
cases).  The traditional distance ladder starting with star clusters
within the Milky Way can be used, as in the HST Key Program value of
$H_0=72\pm 8\kmsmpc$ (\citejap{Freedman2001}), or a more accurate value
obtained by absolute calibration of the Cepheid distance scale using
the maser galaxy NGC4258, yielding $H_0=73.8\pm 2.4\kmsmpc$ (\citejap{Riess2011}).

\begin{figure}[ht]
\center{\includegraphics[scale=0.55]{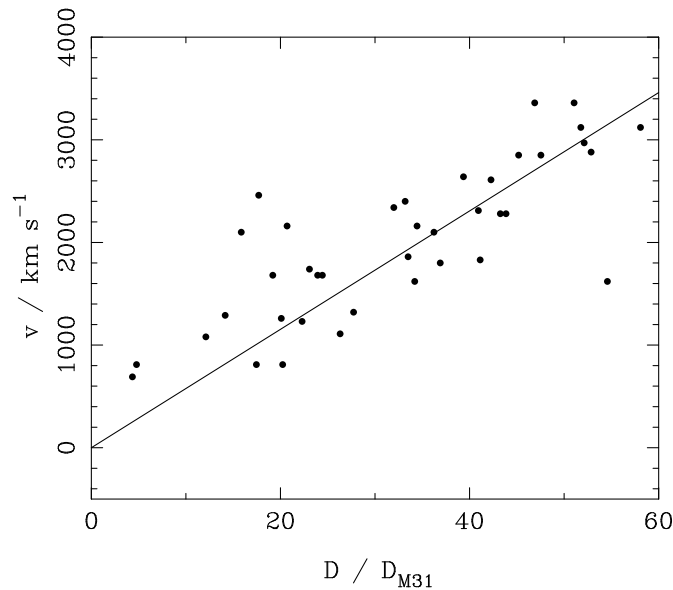}}
\caption{Data on nearby SNe Ia (taken from the compilation of \citejap{Tonry2003}) 
give accurate enough distances that 
we can see clearly the dispersion in the $D-z$ relation caused
by peculiar velocities. This is sporadic: we can have lucky
regions where the
dispersion is low, and others, such as around $D/D_{\rm M31}=20$, 
where it blows up. Typically,
we can see that high-precision distances to perhaps
$D/D_{\rm M31}=50$ would be required for a convincing demonstration of an
underlying linear relation. Again, the solid line
shows the modern truth, assuming
$H_0=73\kmsmpc$ and a distance to M31 of 0.79 Mpc.}\label{peacockfig03}
\end{figure}

Supernovae are especially useful in studying the expansion at larger
distances, since they can readily be detected to $z\simeq1$
(or beyond with effort) -- hence the ability of the SNe Hubble diagram
to probe cosmic acceleration. Figure \ref{peacockfig03}
shows the SNe Hubble relation
in Lundmark's form out to $D=60D_{\rm M31}$, where we can see that
the relation has quiet and noisy regions: the deviation from
uniform expansion is episodic. If we had data only at
$D<30D_{\rm M31}$, there would hardly be evidence for any
correlation between distance and redshift, much less a linear
relation. Things only improve when we probe to 40 or
50 times $D_{\rm M31}$.

\begin{figure}[ht]
\center{\includegraphics[scale=0.50]{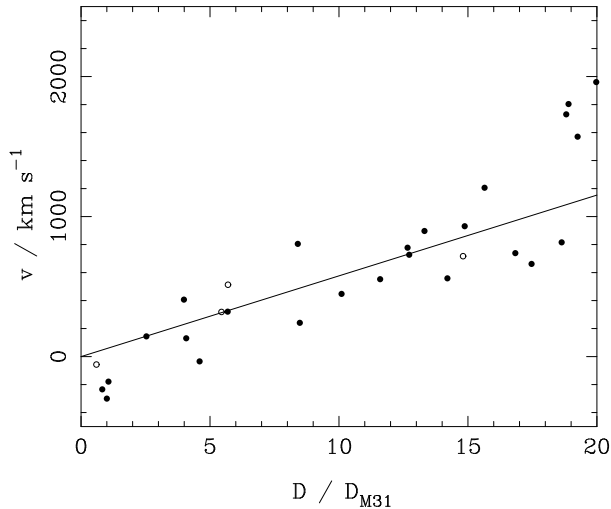}}
\caption{The local Hubble flow is `colder' than in typical regions,
so in fact a linear $D-z$ relation might be detected given
data to $D/D_{\rm M31}\simeq15-20$. This is demonstrated by the 
local Cepheid data (taken from \citejap{Freedman2001}).
Again, the solid line shows the modern truth, assuming
$H_0=73\kmsmpc$ and a distance to M31 of 0.79 Mpc.}\label{peacockfig04}
\end{figure}

Once we get close enough that Cepheids can be detected
(20 Mpc or so) they are a better probe than SNe, since they
are simply more numerous while the distance precision is
comparable. Figure \ref{peacockfig04} plots local Cepheid data and shows that,
closer than the noisy region at $D=20D_{\rm M31}$, we are lucky
enough to experience an unusually quiet part of the Hubble
flow (a fact that has puzzled many workers: e.g.
\citejap{Governato1997}). Although the SNe data show that
this is in fact globally unrepresentative, it is clear that
one could be forgiven for claiming a well-defined
linear $D(z)$ relation given results out to
$D=20D_{\rm M31}$ (although not with any great significance
for distance limits twice as small).

\begin{figure}[ht]
\center{\includegraphics[scale=0.50]{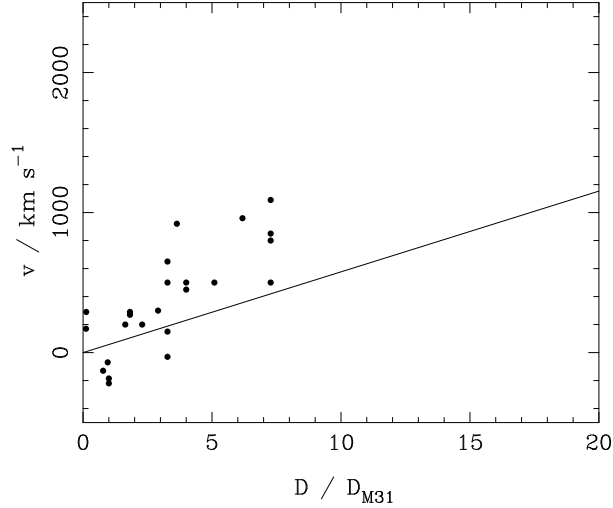}}
\caption{Hubble's 1929 data (largely Slipher's velocities, recall), on the
same scale as the previous plot. Again, the solid line
shows the modern truth, assuming
$H_0=73\kmsmpc$ and a distance to M31 of 0.79 Mpc.
Owing to the distance units as a
ratio, the form of this plot is independent of the assumed
absolute distance calibration, so that the incorrect Cepheid
calibration used by Hubble in deriving $H\simeq 500\kmsmpc$
does not contribute here. Nevertheless, the slope of the relation
is completely wrong, so Hubble's distance estimates were hugely
in error through an independent additional effect.}\label{peacockfig05}
\end{figure}

\section{Hubble's 1929 analysis}

With the above perspective, what are we to make of \citeauthor{Hubble1929}'s
1929 paper, in which a relation between distance and redshift
was announced?
Hubble used a sample of 24 nebulae, 20 of which 
had redshifts measured by Slipher; and with a maximum redshift of
$1100\kms$ the sample is no deeper than that available to Slipher
in 1917. One might therefore have expected that the mean redshift
after dipole subtraction would not be significantly non-zero.
But treating Hubble's sample in the same way as Slipher's reduces the mean
redshift from $373\kms$ to $197\kms$ with an rms of $343\kms$ -- which
is a $2.8\sigma$ deviation from zero. 
The extra significance is aided by the inclusion of the LMC and
SMC: being Southern objects, they constrain the dipole velocity more
strongly and prevent solutions in which the mean redshift is made
as low as is the case with Slipher's sky coverage. Since the Magellanic systems
are so nearby as to be almost part of the Milky Way, the case for including
them is not obvious. 

Let us now see what is added by Hubble's distance data.
His greatest distance was only the
rather modest $D=7.5D_{\rm M31}$, so one might have expected no significant
claims: we have seen from modern data that a linear distance-redshift
relation would not reveal itself clearly with data of twice
this depth. But in fact Hubble's data
do exhibit a correlation between
distance and redshift (Figure \ref{peacockfig05}).
This plot is made in the Lundmark
form used earlier, showing distance as $D/D_{\rm M31}$; and
comparing with the
line representing modern `truth' we see that the slope is
completely wrong. It should be clearly noted that this is not
the same phenomenon as Hubble's overestimation
of $H_0$, because we have used different coordinates. Plotting distance
as $D/D_{\rm M31}$ should remove any calibration errors; thus
Hubble's distances suffer from an entirely distinct additional problem of
internal inconsistency, in addition to the well-known
miscalibration of his Cepheid scale. The symptom
is effectively that the distances for all the most distant objects
are strongly underestimated. This could be suggestive of Malmquist
bias: the distances presented by Hubble go well beyond what
was possible with Cepheids in those days, so Hubble had switched
to using the brightest individual stars as standard candles.
These have a substantial dispersion, so the most distant
galaxies for which such distances can be inferred will be those
where individual stars are abnormally luminous -- causing the
distances to be underestimated. The effect is however extremely
large (roughly a factor 2 in distance), and a simpler alternative explanation is
that Hubble may have simply mistaken compact HII regions in the more
distant galaxies for individual stars (\citejap{Sandage1958}). This is
undeniably a great irony: by combining Slipher's effectively perfect
velocity data with distance estimates that are so badly flawed,
Hubble nevertheless routinely receives sole credit for the
discovery of the expanding universe (including the 
assertion that he measured the redshifts, which is frequently
encountered in popular accounts -- and too often even in those
written by professional scientists).

\begin{table}
\caption{Statistics of various early redshift samples, showing the
influence of correction for Solar motion. This is quoted in Cartesian
components within a J2000 coordinate system, in units of $\kms$.
The dipole is the least-square fit to a dipole-only model; but nevertheless
the mean residual ($\langle v\rangle$) can be significantly positive, when the population
standard deviation ($\sigma_v$) is converted to a standard error ($\sigma_v/N^{1/2}$).
In the case of \citet{Hubble1929} we show results with the full sample and also
excluding the LMC/SMC.
}
\begin{center}
\begin{tabular}{| c | r | r r r | r r | c |} \hline
Sample & $N$ & $v^\odot_x$ & $v^\odot_y$ &$v^\odot_z$ & $\langle v \rangle$ & $\sigma_v$ & Significance \\ \hline 
S17 & 25 & 0 & 0 & 0 & 502 & 422 & $5.9\sigma$ \\
S17 & 25 & 566 & $-356$ & $-268$ & 145 & 414 & $1.8\sigma$ \\
S23 & 41 & 0 & 0 & 0 & 571 & 439 & $8.3\sigma$ \\
S23 & 41 & 467 & $-856$ & $-298$ & 201 & 508 & $2.5\sigma$ \\
H29 & 24 & 0 & 0 & 0 & 373 & 371 & $4.9\sigma$ \\
H29 & 24 & 462 & $-317$ & $-117$ & 197 & 343 & $2.8\sigma$ \\
H29 & 22 & 0 & 0 & 0 & 386 & 385 & $4.7\sigma$ \\
H29 & 22 & 426 & $-205$ & $-200$ & 159 & 370 & $2.0\sigma$ \\ \hline
\end{tabular}
\end{center}
\end{table}

\begin{table}
\caption{The equivalent of Table 1, but now assuming a model
of an explicit non-zero (constant) mean velocity. Note
that, with respect to the fits of Table 1, the best-fitting
dipole is different and the dispersion
in the residuals is smaller -- representing a more
significant detection of a non-zero mean.
}
\begin{center}
\begin{tabular}{| c | r | r r r | r r | c |} \hline
Sample & $N$ & $v^\odot_x$ & $v^\odot_y$ &$v^\odot_z$ & $\langle v \rangle$ & $\sigma_v$ & Significance \\ \hline
S17 & 25 & 246 & $25$ & $-430$ & 566 & 328 & $8.6\sigma$ \\
S23 & 41 & 81 & $-109$ & $661$ & 805 & 364 & $14.1\sigma$ \\
H29 & 24 & 323 & $-267$ & 113 & 315 & 306 & $5.0\sigma$ \\
H29 & 22 & 322 & $-483$ & 424 & 493 & 286 & $8.1\sigma$ \\ \hline
\end{tabular}
\end{center}
\end{table}

\begin{table}
\caption{Statistics of Hubble's 1929 data. The population standard deviation, $\sigma_v$,
about the best-fitting linear model is given with and without dipole correction.
Results are given with the full sample and also
excluding the LMC/SMC. The units of $H$ are $\kms$, since distances are in
units of $D_{\rm M31}$.
}
\begin{center}
\begin{tabular}{| r | r r r | r r |} \hline
$N$ & $v^\odot_x$ & $v^\odot_y$ &$v^\odot_z$ & $H$ & $\sigma_v$  \\ \hline
24 & 0 & 0 & 0 & 373 & 371 \\
24 & 67 & $-219$ & 189 & 462 & 192 \\
22 & 0 & 0 & 0 & 386 & 385 \\
22 & 68 & $-235$ & 205 & 467 & 199 \\ \hline
\end{tabular}
\end{center}
\end{table}

Another interesting aspect of Hubble's analysis is that he assumes from
the start a model that includes a linear $D(z)$ relation as well as a reflex dipole:
\be
v = HD - {\bf v}_\odot \cdot {\bf\hat r}.
\ee
The famous $v-D$ plot from his 1929 paper shows not the
raw velocities, but rather the velocities corrected by the
dipole that best-fits the above relation -- i.e. the plot
has been manipulated in order to make a linear relation
look as good as possible. Admittedly, Hubble does state that
the data ``\dots indicate a linear correlation between distances 
and velocities, whether the latter are used directly or corrected for solar motion.'',
but we are not shown the uncorrected plot -- and we have seen
in the case of Slipher's data that the Solar motion can change
the picture very substantially.

It is therefore worth looking carefully at the statistics of the
various samples that have been discussed, and these are collected
in Table 1. From this, it is apparent that Hubble was a little
fortunate with his
1929 data: the mean redshift
after dipole correction is substantially more significant
than Slipher's 1917 results -- and more so than even
Slipher's much deeper data of 1923. But this ceases to be
true when the LMC and SMC are removed from the 1929 sample.
Hubble's sample is therefore poised to deliver evidence
for an expanding universe, even before adding distance data.

Because of these non-zero mean velocities, we should
make it clear that
there are (at least) three distinct models worth considering:
\be
\eqalign{
(1)\ v &= - {\bf v}_\odot \cdot {\bf\hat r} \cr
(2)\ v &= \bar v - {\bf v}_\odot \cdot {\bf\hat r} \cr
(3)\ v &= HD - {\bf v}_\odot \cdot {\bf\hat r}.
}
\ee
Model 1 is all that has been considered so far.
\citet{Hubble1929} quotes $H=513\pm 60\kmsmpc$, which represents an
$8.6\sigma$ detection of a linear $D(z)$ (model 3) in comparison
to model 1 (pure dipole). But this is not the right question, since
we have seen that the mean redshift in Hubble's data is clearly
non-zero. Since model (3) naturally predicts a non-zero mean,
we have to ask whether the distance data add anything
significant beyond this fact. There are a number of ways in
which this can be assessed, and the simplest is to look at
the size of the residuals about the best-fitting linear+dipole
model. Table 2 shows these results, again with and without
the Magellanic systems. 

The striking feature of Table 2 is that the standard deviations 
in the residuals are smaller than in Table 1. This may seem
puzzling at first sight, since in each case the only correction
made to the data has been to remove a dipole. But in model
1 (which is what was used by \citejap{Slipher1917}), we are attempting
to minimise the mean square velocity, not the dispersion about
the mean; this naturally pushes the mean low. 
If we are open to the possibility of a non-zero
mean, then we need to minimise the standard deviation. This
yields a different dipole and a more secure detection of
the mean. Indeed, the remarkable conclusion of Table 2
is that Slipher's data alone provide a very secure
detection of a non-zero mean velocity: $8.6\sigma$ in
1917 and $14.1\sigma$ in 1923. This significance is
slightly overestimated because of the reduction in 
degrees of freedom caused by best-fitting the dipole.
But this would only be a slight effect -- especially with
the $N=41$ of 1923. This huge significance vindicates Eddington's ready
acceptance of a non-zero mean velocity without the need
for a detailed analysis.

We now consider the fits of model 3, which are given in Table 3.
Model 3 is clearly a better fit than model
2: for Hubble's full sample, the rms residual is reduced from
$306\kms$ to $192\kms$ -- but is this reduction significant?
The question is whether the low rms might be simply a
statistical fluctuation downwards from a true value of around 300. 
For Gaussian distributions, the standard deviation, $\epsilon$, on the
estimate of the population standard deviation, $s$, is 
$\epsilon = s/[2(N-1)]^{1/2}$. We are therefore
comparing $192\pm28$ with $306\pm 45$, which is only
a $2.1\sigma$ difference. But this
statement applies only to independent samples, whereas we
have the same data fitted with two different models. 
A better way to deal with this objection, plus the issue that the
dipole is fitted to the data, is to use
Monte Carlo: we hypothesise that there is no information
in Hubble's distances, so we randomly permute them, and fit
in each case a model of linear $D(z)$ plus dipole. This allows us to compute
how often the rms is lowered by as much as is observed relative
to model 2.
The answer is about 1 in 23,000 for Hubble's full sample
(a $3.9\sigma$ deviation), or 1 in 3000 if we ignore 
the LMC/SMC (a $3.5\sigma$ deviation). Thus the distance
estimates do contain evidence for a correlation between
distance and redshift -- but at a lesser additional degree of
significance than the basic fact that the mean redshift
tends to be positive.

The picture that emerges from this study is thus that Hubble's 1929
work was perhaps more an exercise in
validation of a linear $D(z)$ than a discovery.
Hubble's closing quote that ``\dots the velocity-distance
relation may represent the de Sitter effect\dots'' shows
that he was certainly aware of the theoretical prediction
that motivated earlier studies, such as that of \citet{Lundmark1924}.
Hubble is not explicit in his introduction about the role
that theory played in his work, although he did state that previous
(un-named) investigations had sought
``\dots a correlation between apparent radial velocities
and distances, but so far the results have not been
convincing''. Since this previous work was motivated
by a search for the de Sitter effect, we can conclude
that Hubble was influenced by the same theoretical
prior as Lundmark in 1924 -- and it is debatable which
of these investigations achieved greater success in
tracking down their quarry.

\section{Peculiar velocities today}

\subsection{Velocities and structure formation}

Slipher's demonstration that the Milky Way is not at rest is as
revolutionary a moment as Bradley's proof in 1728 from
stellar aberration that the Earth is in motion. We see
this effect today most clearly in the dipole component of the
Cosmic Microwave Background, 
which measures the Solar motion as $368\kms$. The fact
that this differs from Slipher's $700\kms$ is further proof that
his sample of galaxies is not deep enough to be a fair sample
of the universe, from which one could really expect to
measure the expansion.

But Slipher's data were deep enough to
show that all galaxies have a random component
to their velocities, so that the universe contains a
peculiar velocity field. These deviations from the general
expansion have been of great importance in
cosmological research over the past several decades. The
significance of peculiar velocities is that they must have
their origin in the gravitational forces that cause the
growth of cosmic inhomogeneities. If the dimensionless
density fluctuation, $\delta$ is defined by
$\rho = (1+\delta)\langle\rho\rangle$, then conservation
of mass requires
\be
{\partial\delta\over\partial t} = -{\bf \nabla\cdot (1+\delta)u}
\simeq -{\bf \nabla\cdot u},
\ee
where $u$ is a comoving peculiar velocity: the physical peculiar
velocity is $\delta {\bf v} = a {\bf u}$, where $a(t)$ is the
dimensionless cosmic scale factor. The last equality holds in
the linear limit of small density fluctuations.

The perturbation growth rate is more commonly written in terms of the
logarithmic derivative, $f_g$:
\be
f_g\equiv{\partial \ln\delta\over \partial \ln a} =
{1\over H\delta} {\partial\delta\over\partial t} = -{1\over H\delta}{\bf \nabla\cdot u}.
\ee
In this form, the growth rate depends purely on the density of the
universe, and in years gone by this was seen as a powerful
route towards measuring the matter density. Today, with the density
measured accurately via the CMB, the focus
has shifted to using the growth of structure as a test of Einstein's
relativistic theory of gravity. This boils down to the common approximation
\be
f_g \simeq \Omega_m(a)^\gamma
\ee
where $\gamma\simeq 0.55$ for Einstein gravity, largely
independent of the value of any cosmological constant, but
non-standard gravity models can yield values that differ
from this by several tenths
(\citejap{Peebles1980}; \citejap{Lahav1991}; \citejap{Linder2007}).

The motivation for thinking about deviations from Einstein gravity is not simply
that it is always a good idea to verify fundamental
assumptions of a field where possible. The possibility that
Einstein's theory may be incorrect derives its motivation from
the most radical constituent of modern cosmology: the deduction
that roughly 75\% of the mean density is contributed by a
nearly uniform component termed dark energy. So far, the
properties of this substance are empirically indistinguishable
from a cosmological constant or vacuum energy, but are we really
sure that the dark energy exists? The doubt comes not through
uncertain data, but because the inference derives entirely from
the expansion history of the universe, which is interpreted
via the Friedmann equation
\be
H^2(a) = H_0^2\left(\Omega_r a^{-4} + \Omega_m a^{-3}
+ (1-\Omega_{\rm total})a^{-2} + \Omega_v \right).
\ee
Empirically, it is impossible to match the data on $H(a)$
using only known matter constituents without adding a constant term on the 
right-hand side.
But this might simply say that the Friedmann equation is
wrong; it could be that some alternative to Einstein gravity
might generate a Friedmann equation containing a constant term
without needing to introduce dark energy as a physical
substance. The way to distinguish between these options is to look
for a scale dependence of any gravitational modifications,
and the peculiar velocities associated with the
growth of structure are a perfect tool for this job,
since they measure the strength of gravity on scales
of $\sim 10 - 100 \mpcoh$. As a result studies of the
growth rate of perturbations have, together with gravitational
lensing, assumed huge importance in recent years as an
industry has built up around cosmological tests of
gravity (see e.g. \citejap{Jain2010}).

\subsection{Direct velocity measurements}

There are two main ways in which the growth rate can be
measured, and the first to receive attention was the most
direct: estimate the peculiar velocity field from data.
To do this requires some means of estimating distances, since
\be
\delta v = v -  HD
\ee
(assuming low enough redshifts that the cosmological and
Doppler peculiar redshifts simply add; at higher redshifts
we should multiply the $1+z$ factors). Taking the divergence
of the peculiar velocities inferred in this way is problematic 
since we only observe the radial component. This can be cured by
adding the assumption that density perturbations under
gravitational instability are expected to be in the growing
mode, in which the velocities are irrotational.
Thus ${\bf u = -\nabla\psi}$, where $\psi$ is a velocity
potential -- which can be measured by integrating along the
line of sight (\citejap{Bertschinger1989}).

There are two problems with this method. The difficulty of
principle is that the divergence of ${\bf u}$ is proportional
to $\delta$ times $f_g$, so we need to know the absolute
level of density fluctuations. This is not so easy when using
galaxies as tracers, because they are {\it biased\/}:
$\delta_{\rm gal}\simeq b\delta$ on large scales. Thus
we measure not $\Omega_m^\gamma$, but $\Omega_m^\gamma/b$.

The second difficulty is the practical one: the only tracers
of peculiar velocities that have high space densities are galaxies,
so we need to treat them as some kind of standard candles
in order to deduce distances. Even with luminosities calibrated
by an internal velocity (the `Tully-Fisher method' for spirals;
the `fundamental plane' for ellipticals), the distances are
good to only around 20\%, and this scatter necessitates
careful statistical treatment in order to avoid Malmquist bias
and related effects.

A number of studies appeared in the 1990s claiming that these
problems could be cured (e.g. \citejap{Sigad1998}), and the consistent
result was a high value of $\Omega_m^\gamma/b$, close to unity.
It was possible to argue from the statistics of the collapse of
dark-matter haloes that $b$ should never be very much less
than unity for any given class of galaxy (e.g. \citejap{Cole1989};
\citejap{Mo1996}), and therefore these results were seen as
supporting a high matter density -- most naturally $\Omega_m=1$.
This flat model was known to be in good agreement with the
early limits on CMB fluctuations; these ruled out low-density
open models, so that a cosmological constant was the only option
if a low matter density $\Omega_m\simeq 0.2$ was preferred. A
good body of evidence existed at that time (ranging from large-scale galaxy
clustering to the baryon fraction in rich clusters) to suggest
that $\Omega_m=1$ was too high, so there were strong arguments
in favour of a $\Lambda$-dominated model (\citejap{Efstathiou1990});
the last resistance to such a model crumbled with the
arrival of the high-redshift supernova data in the late 1990s.

From this point on, direct use of peculiar velocity estimates has
been somewhat neglected. No convincing explanation has really been given
for why the 1990s velocity measurements gave what is now
considered to be too high a density, and indeed discrepant
results continue to exist in the form of apparent `streaming
velocities' that are inconsistent with what we think we know
about the mass distribution (e.g. \citejap{Watkins2009};
\citejap{Kashlinsky2009}). But this is probably a common
situation in science: where the evidence for a standard model is
strong, discrepant results are most likely to be flawed and so
a community is rightly reluctant to invest too much effort in
understanding what has gone wrong. Sometimes this approach
will ignore the key to a revolution, of course, but the large
dispersion in individual peculiar velocity estimates means that
a claimed rejection of $\Lambda$CDM based on such data will
continue to be treated skeptically.

\subsection{Redshift-space distortions}

Nevertheless, peculiar velocities remain a major tool in
conventional cosmology. This is because of the existence
of major galaxy redshift surveys, where up to $10^6$ galaxies
are used to build up a picture of the 3D distribution of
luminous matter. Such surveys have turned out to be
fantastic statistical tools, because the power spectrum of
fluctuations contains characteristic lengths that can be
measured and used as a diagnostic of conditions in the
early universe. Chief among these are the relatively broad
curvature in the spectrum around the horizon size at
matter-radiation equality, and the sharper feature at the
acoustic horizon following last scattering. These are
mainly sensitive to the density of the universe, and gave some
of the first evidence for low-density models, as mentioned above.
Today, the frontier is to measure the angular
scale corresponding to these lengths as a function
of redshift, mapping the $D(z)$ relation with standard rulers.

But the 3D picture given by redshift surveys is distorted
in the radial direction by peculiar velocities, and in a
complicated way that is correlated with the actual structures
to be studied. Rather than being a bug, this is a feature:
it causes observed galaxy clustering to be anisotropic in
a way that allows a very precise statistical characterization of
the amplitude of peculiar velocities.

\begin{figure}[ht]
\center{\includegraphics[scale=0.65]{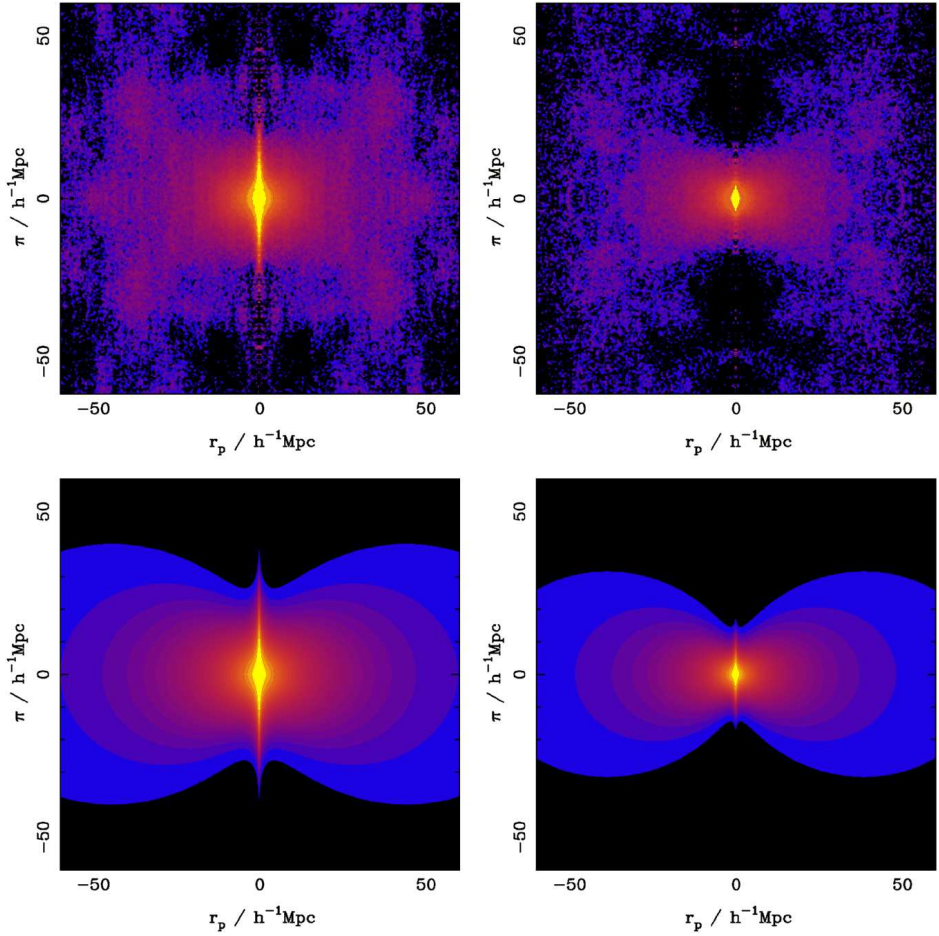}}
\caption{Redshift-space clustering as measured in the GAMA survey,
split by colour of galaxy, together with theoretical models
allowing for different degrees of linear flattening and of
the pairwise dispersion (red galaxies to the left; blue to
the right). As expected, the higher-mass haloes hosting red
galaxies give them a higher pairwise dispersion and a higher
bias -- thus a lower large-scale flattening.}\label{peacockfig06}
\end{figure}

Redshift-space distortions of clustering were first given a comprehensive
analysis by \citet{Kaiser1987}. In the limit of a distant observer, where all
pairs subtend small angles, the apparent anisotropic power spectrum
for some biased tracer is given in linear theory by
\be
\eqalign{
P(k,\mu) &= P_m(k) (b+f_g\mu^2)^2 \cr
&= b^2P_m(k) (1+\beta \mu^2)^2; \quad \beta\equiv f_g/b,\cr}
\ee
where $P_m(k)$ is the linear matter power spectrum,
$f_g$ is the desired growth factor, $f_g\equiv d\ln\delta/d\ln a$ and
$b$ is a linear bias parameter.
It is common to find this model extended to allow for `Fingers of God',
in which the density field is convolved radially by random virialized velocities in haloes.
Most usually an exponential pairwise velocity distribution is adopted,
with rms $\sigma_p$ (expressed as a length),
leading to Lorenzian line-of-sight damping in Fourier space:
\be
P_s(k,\mu)= b^2P(k) (1+\beta \mu^2)^2 / (1+k^2\mu^2\sigma_p^2/2).
\ee
The original derivation is only valid for small density fluctuations
in the linear regime, but this expression has been used with some success
inserting the non-linear real-space power spectrum of galaxies in
place of $b^2P(k)$. 

An example of such modelling is shown in
Figure \ref{peacockfig06}, which presents preliminary results from the GAMA
survey (\citejap{Driver2011}). Here we see the galaxy population
split by colour, with the result that the red population shows
larger fingers of God, and less pronounced large-scale flattening
(a smaller value of $\beta$). Both these results can be understood
in terms of the typical mass of the dark-matter haloes hosting the
galaxies: where this is larger, the small-scale velocity dispersion
is larger and the large-scale clustering amplitude increases
(which reduced $\beta$, since it is $\propto 1/b$).

\begin{figure}[ht]
\center{\includegraphics[scale=0.5]{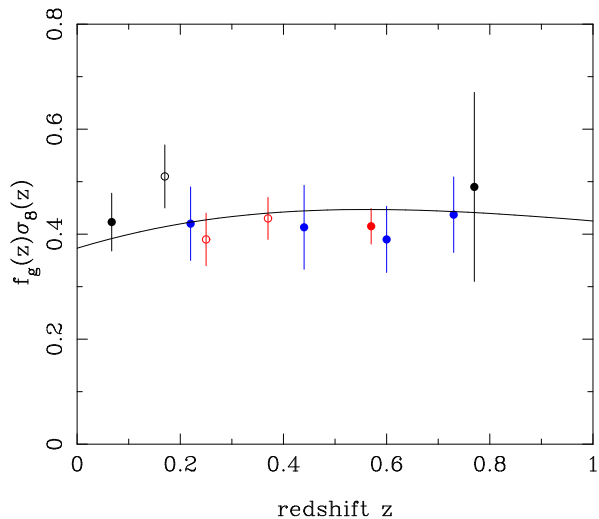}}
\caption{A plot of various measurements of the linear growth of
density perturbations inferred from redshift-space distortions
(see e.g. \citejap{Beutler2012} for a compilation of the data).
The solid line shows a default flat $\Lambda$-dominated model
with $\Omega_m=0.25$ and $\sigma_8=0.8$, which matches the
data very well (perhaps too well: 9 out of 10 measurements
agree with the theory to within 1$\sigma$).}\label{peacockfig07}
\end{figure}

The bias parameter is hard to predict a priori,
meaning that this method is unable to yield a direct measurement
of the perturbation growth rate without additional assumptions.
The way this is
dealt with in practice is to realise that the real-space
clustering amplitude of galaxies is observable, so that the
bias can be measured if a model for the mass fluctuations
is assumed. At the level of a consistency check, this can be
a standard $\Lambda$CDM model taken from CMB and other data.
A slightly more general way of putting this is to say that galaxy
data determine $b\sigma_8$, where $\sigma_8$ is the usual
normalization measure of density fluctuations: the linear-theory
extrapolated fractional rms variation when averaged in spheres
of radius $8\mpcoh$. Thus the slightly unlovely combination
$f_g(z)\sigma_8(z)$ can be measured in an approximately
model-independent fashion. A compilation of recent 
estimates of this quantity is shown in Figure \ref{peacockfig07}, which
shows impressive consistency with the standard model,
indicating that Einstein's relativistic theory of gravity
can be verified at about the 5-10\% level on scales
$\sim 10-30\mpcoh$ over a wide range of cosmological time.
This is hardly yet at the level of precision of solar-system
tests, but this limit will in due course be brought down to the
per cent level by future experiments such as ESA's Euclid
satellite. This should be launched around 2020 (\citejap{Laureijs2011}), and
will provide redshifts for around 50 million galaxies in a redshift band around
$z\simeq 2$, as opposed to current studies which are based on
in total around one million galaxies in the smaller volume at
$z<1$.

\section{Discussion and conclusions}

It is irresistible to speculate about how Slipher might feel
were he able to hear us talk calmly of having measured a
million galaxy redshifts, and how we plan to increase
this number a hundredfold -- when each of his measurements
cost him several nights standing alone in the cold. But
from the anecdotes aired at this meeting, one suspects
he might not have been all that jealous, since he seems
to have found a deep attraction in the basic process of
observing. And it is undeniable that something is in danger
of being lost as we pursue large-scale cosmology with an
industrial efficiency; there are declining opportunities
for young astronomers to work at telescopes and experience
that sense of a mystical connection to the cosmos that
comes from standing by a telescope in the dark under
a clear sky. As the machines become larger, one way
of retaining that sense of wonder is to remember the
efforts of the pioneers.

And Slipher was a great pioneer; not simply through his instrumental
virtuosity in achieving reliable velocities where others had failed,
but through the clarity of reasoning he applied. Respect for what he
did and did not claim can only be increased by the exercise presented
here of analyzing his data as if the information was freshly available,
and trying as best we can to rid the mind of modern preconceptions. At the depth
to which he worked, and with the restrictions of sky coverage, it was
hard for the signature of a general expansion to stand out.
But it is hard not to wonder
what would have happened if data for more southerly or even slightly more distant
galaxies had been available; Slipher's comment in 1917 that
generally positive velocities ``\dots might suggest that the spiral
nebulae are scattering\dots'' suggests that he would have been open to
the conclusion of a general expansion. As shown above, such a result
can actually be obtained from Slipher's 1917 data
(a $>$$8\sigma$ detection of a non-zero mean velocity, even after
allowance for the best-fitting Solar dipole), and the signal
rises to $14\sigma$ with the expanded dataset that Slipher
gave freely to Eddington and others in 1923. It is more than a little
surprising that no-one attempted to repeat Slipher's 1917 work
with this expanded material, in which case Slipher could have
been clearly established as the discoverer of the expanding universe.

Unlike Hubble and other
workers from the 1920s, Slipher in 1917 lacked the theoretical
prior of a predicted linear distance-redshift relation, which de Sitter
only published the same year. Slipher was simply looking for a message
that emerged directly from the data, and it is therefore all the more
impressive that he was able to reach his beautiful 1917 conclusions
concerning the motion of the Milky Way and the nature of spiral nebulae as similar
stellar systems. But this is characteristic of Slipher's work: right from
his early assertion that the velocity of M31 must be Doppler in origin,
he was willing to stick his neck out and state firm conclusions when
he believed that these were supported by the data.
Rather than using hindsight to regret that he did not focus
on the non-zero mean velocity of his data, we should look on
with admiration at how much he was nevertheless able to learn from the 
observations he had gathered.

By adding distance data to existing velocities, \citet{Hubble1929} claimed
not only that the mean velocity was a redshift, but that redshift correlated
linearly with distance. We have seen that Hubble was fortunate in a number
of ways to have been able to make such a claim with the material to hand:
(1) peculiar velocities are unusually low in the local
volume; (2) his mean redshift was higher than Slipher's in 1917,
despite the sample containing no greater velocities;
(3) he included the LMC and SMC, which could be viewed
as unjustified; (4) his distance estimates were flawed
in two distinct ways. Also, Hubble considered from the
outset only the hypothesis of a linear relation between
distance and redshift, and never asked how much his
information added to the simple statement that the mean
velocity was positive (which we have seen accounts for
the majority of the statistical weight in his result).
Hubble admitted that he was following up previous
searches for a distance-redshift correlation, and
these studies were explicitly motivated by the theoretical
prior of the de Sitter effect. If this prediction had
been absent in 1929, one wonders if claims of a
linear distance-redshift relation would have been made at 
that time.

If the data in 1929 were really too shallow for a truly robust proof
of a linear distance-redshift relation, when was this first
seen unequivocally? Credit is often given to \citet{Hubble1931},
who pushed the maximum velocity out to $20,000\kms$ -- ten times
what had been achieved by Slipher. But the distances used in that
paper were based on the same unjustified 
assumption used by Lundmark in 1924: that
galaxies could be treated as standard objects. 
Indeed, Hubble gives a pre-echo of this argument in his
1929 paper, referring to the large redshift of NGC7619.
Because galaxies at these distances lacked any sort
of well-justified distance estimates,
one could imagine that the 1931 paper should have received a
good deal of critical skepticism
-- but by this time a linear $D(z)$ was already regarded as 
having been proved.

In fact, right through the 1980s,
cosmology journals and conferences were treated to a continuing
critique of a linear $D(z)$ as deduced from galaxy data
by Irving Segal (e.g. \citejap{Segal1989}). Segal made major contributions to quantum
field theory, and could hardly be dismissed as a crank; the
basic problem is that, even when calibrated dynamically
as in the Tully-Fisher method, the scatter in galaxy properties
is so large that getting distances to better than around 20\%
is not feasible. Thus it was only really in the
1990s, with HST extending the reach of Cepheids and SNe Ia giving
accurate distances, that we could verify what had been generally
assumed to be true since 1929/1931.

But if the work on the distance scale in the 1990s closed the chapter
on the local distance-redshift relation that was begun in the 1920s,
Slipher's other main legacy to modern cosmology remains as relevant
as ever. The peculiar velocity field that he discovered has become
one of the centrepieces of modern efforts to measure the nature of
gravity on cosmological scales. Hence we have come full circle, from
assuming the correctness of Einstein's relativistic gravity (and of the de
Sitter solution in particular) to search for
evidence of expansion in the 1920s, to the present-day use of data on peculiar velocities
to tell us if the theory is correct. Slipher would probably have
been happy to see things being done in this direction.


\bibliography{lowell_jap_final}

\begin{thebibliography}{}
\expandafter\ifx\csname natexlab\endcsname\relax\def\natexlab#1{#1}\fi
\expandafter\ifx\csname url\endcsname\relax
  \def\url#1{\texttt{#1}}\fi
\expandafter\ifx\csname urlprefix\endcsname\relax\def\urlprefix{URL }\fi
\providecommand{\eprint}[2][]{\url{#2}}

\bibitem[{{Bernstein} \& {Feinberg}(1986)}]{Bernstein1986}
{Bernstein}, J., \& {Feinberg}, G. (eds.) 1986, {Cosmological constants: Papers
  in modern cosmology} (Columbia University Press)

\bibitem[{{Bertschinger} \& {Dekel}(1989)}]{Bertschinger1989}
{Bertschinger}, E., \& {Dekel}, A. 1989, {Recovering the full velocity and
  density fields from large-scale redshift-distance samples}, \apjl, 336, L5

\bibitem[{{Beutler} et~al.(2012){Beutler}, {Blake}, {Colless}, {Jones},
  {Staveley-Smith}, {Poole}, {Campbell}, {Parker}, {Saunders}, \&
  {Watson}}]{Beutler2012}
{Beutler}, F., {Blake}, C., {Colless}, M., {Jones}, D.~H., {Staveley-Smith},
  L., {Poole}, G.~B., {Campbell}, L., {Parker}, Q., {Saunders}, W., \&
  {Watson}, F. 2012, {The 6dF Galaxy Survey: $z\simeq0$ measurements of the
  growth rate and {$\sigma$}$_{8}$}, \mnras, 423, 3430

\bibitem[{{Cole} \& {Kaiser}(1989)}]{Cole1989}
{Cole}, S., \& {Kaiser}, N. 1989, {Biased clustering in the cold dark matter
  cosmogony}, \mnras, 237, 1127

\bibitem[{{de Sitter}(1917)}]{deSitter1917}
{de Sitter}, W. 1917, {Einstein's theory of gravitation and its astronomical
  consequences. Third paper}, \mnras, 78, 3

\bibitem[{{Driver} et~al.(2011)}]{Driver2011}
{Driver}, S.~P., et~al. 2011, {Galaxy and Mass Assembly (GAMA): survey
  diagnostics and core data release}, \mnras, 413, 971

\bibitem[{{Eddington}(1923)}]{Eddington1923}
{Eddington}, A.~S. 1923, {The mathematical theory of relativity} (Cambridge
  University Press)

\bibitem[{{Efstathiou} et~al.(1990){Efstathiou}, {Sutherland}, \&
  {Maddox}}]{Efstathiou1990}
{Efstathiou}, G., {Sutherland}, W.~J., \& {Maddox}, S.~J. 1990, {The
  cosmological constant and cold dark matter}, \nat, 348, 705

\bibitem[{{Einstein}(1917)}]{Einstein1917}
{Einstein}, A. 1917, {Kosmologische Betrachtungen zur allgemeinen
  Relativit{\"a}tstheorie}, Sitzungsberichte der K{\"o}niglich Preu{\ss}ischen
  Akademie der Wissenschaften (Berlin), Seite 142-152., 142

\bibitem[{{Freedman} et~al.(2001){Freedman}, {Madore}, {Gibson}, {Ferrarese},
  {Kelson}, {Sakai}, {Mould}, {Kennicutt}, {Ford}, {Graham}, {Huchra},
  {Hughes}, {Illingworth}, {Macri}, \& {Stetson}}]{Freedman2001}
{Freedman}, W.~L., {Madore}, B.~F., {Gibson}, B.~K., {Ferrarese}, L., {Kelson},
  D.~D., {Sakai}, S., {Mould}, J.~R., {Kennicutt}, R.~C., Jr., {Ford}, H.~C.,
  {Graham}, J.~A., {Huchra}, J.~P., {Hughes}, S.~M.~G., {Illingworth}, G.~D.,
  {Macri}, L.~M., \& {Stetson}, P.~B. 2001, {Final Results from the Hubble
  Space Telescope Key Project to Measure the Hubble Constant}, \apj, 553, 47

\bibitem[{{Governato} et~al.(1997){Governato}, {Moore}, {Cen}, {Stadel},
  {Lake}, \& {Quinn}}]{Governato1997}
{Governato}, F., {Moore}, B., {Cen}, R., {Stadel}, J., {Lake}, G., \& {Quinn},
  T. 1997, {The Local Group as a test of cosmological models}, New Astronomy,
  2, 91

\bibitem[{{Hubble} \& {Humason}(1931)}]{Hubble1931}
{Hubble}, E., \& {Humason}, M.~L. 1931, {The Velocity-Distance Relation among
  Extra-Galactic Nebulae}, \apj, 74, 43

\bibitem[{{Hubble}(1929)}]{Hubble1929}
{Hubble}, E.~P. 1929, {A Relation between Distance and Radial Velocity among
  Extra-Galactic Nebulae}, Proceedings of the National Academy of Science, 15,
  168

\bibitem[{{Jain} \& {Khoury}(2010)}]{Jain2010}
{Jain}, B., \& {Khoury}, J. 2010, {Cosmological tests of gravity}, Annals of
  Physics, 325, 1479

\bibitem[{{Janiak}(2004)}]{Janiak2004}
{Janiak}, A. 2004, {Isaac Newton: Philosophical Writings} (Cambridge University
  Press)

\bibitem[{{Kaiser}(1987)}]{Kaiser1987}
{Kaiser}, N. 1987, {Clustering in real space and in redshift space}, \mnras,
  227, 1

\bibitem[{{Kashlinsky} et~al.(2009){Kashlinsky}, {Atrio-Barandela}, {Kocevski},
  \& {Ebeling}}]{Kashlinsky2009}
{Kashlinsky}, A., {Atrio-Barandela}, F., {Kocevski}, D., \& {Ebeling}, H. 2009,
  {A Measurement of Large-Scale Peculiar Velocities of Clusters of Galaxies:
  Technical Details}, \apj, 691, 1479

\bibitem[{{Lahav} et~al.(1991){Lahav}, {Lilje}, {Primack}, \&
  {Rees}}]{Lahav1991}
{Lahav}, O., {Lilje}, P.~B., {Primack}, J.~R., \& {Rees}, M.~J. 1991,
  {Dynamical effects of the cosmological constant}, \mnras, 251, 128

\bibitem[{{Laureijs} et~al.(2011){Laureijs}, {Amiaux}, {Arduini},
  {Augu{\`e}res}, {Brinchmann}, {Cole}, {Cropper}, {Dabin}, {Duvet}, {Ealet}
  et~al.}]{Laureijs2011}
{Laureijs}, R., {Amiaux}, J., {Arduini}, S., {Augu{\`e}res}, J.~.,
  {Brinchmann}, J., {Cole}, R., {Cropper}, M., {Dabin}, C., {Duvet}, L.,
  {Ealet}, A., et~al. 2011, {Euclid Definition Study Report}, ArXiv:1110.3193

\bibitem[{{Lema{\^i}tre}(1927)}]{Lemaitre1927}
{Lema{\^i}tre}, G. 1927, {Un Univers homog{\`e}ne de masse constante et de
  rayon croissant rendant compte de la vitesse radiale des n{\'e}buleuses
  extra-galactiques}, Annales de la Societe Scietifique de Bruxelles, 47, 49

\bibitem[{{Lema{\^i}tre}(1931)}]{Lemaitre1931}
--- 1931, {Expansion of the universe, A homogeneous universe of constant mass
  and increasing radius accounting for the radial velocity of extra-galactic
  nebulae}, \mnras, 91, 483

\bibitem[{{Linder} \& {Cahn}(2007)}]{Linder2007}
{Linder}, E.~V., \& {Cahn}, R.~N. 2007, {Parameterized beyond-Einstein growth},
  Astroparticle Physics, 28, 481

\bibitem[{{Lundmark}(1924)}]{Lundmark1924}
{Lundmark}, K. 1924, {The determination of the curvature of space-time in de
  Sitter's world}, \mnras, 84, 747

\bibitem[{{Mo} \& {White}(1996)}]{Mo1996}
{Mo}, H.~J., \& {White}, S.~D.~M. 1996, {An analytic model for the spatial
  clustering of dark matter haloes}, \mnras, 282, 347

\bibitem[{{Paddock}(1916)}]{Paddock1916}
{Paddock}, G.~F. 1916, {The Relation of the System of Stars to the Spiral
  Nebulae}, \pasp, 28, 109

\bibitem[{{Peebles}(1971)}]{Peebles1971}
{Peebles}, P.~J.~E. 1971, {Physical cosmology} (Princeton University Press)

\bibitem[{{Peebles}(1980)}]{Peebles1980}
--- 1980, {The large-scale structure of the universe} (Princeton University
  Press)

\bibitem[{{Riess} et~al.(2011){Riess}, {Macri}, {Casertano}, {Lampeitl},
  {Ferguson}, {Filippenko}, {Jha}, {Li}, \& {Chornock}}]{Riess2011}
{Riess}, A.~G., {Macri}, L., {Casertano}, S., {Lampeitl}, H., {Ferguson},
  H.~C., {Filippenko}, A.~V., {Jha}, S.~W., {Li}, W., \& {Chornock}, R. 2011,
  {A 3\% Solution: Determination of the Hubble Constant with the Hubble Space
  Telescope and Wide Field Camera 3}, \apj, 730, 119

\bibitem[{{Sandage}(1958)}]{Sandage1958}
{Sandage}, A. 1958, {Current Problems in the Extragalactic Distance Scale.},
  \apj, 127, 513

\bibitem[{{Segal}(1989)}]{Segal1989}
{Segal}, I.~E. 1989, {Direct comparison of observed magnitude-redshift
  relations in complete galaxy samples with systematic predictions of
  alternative redshift-distance laws}, \mnras, 237, 17

\bibitem[{{Sigad} et~al.(1998){Sigad}, {Eldar}, {Dekel}, {Strauss}, \&
  {Yahil}}]{Sigad1998}
{Sigad}, Y., {Eldar}, A., {Dekel}, A., {Strauss}, M.~A., \& {Yahil}, A. 1998,
  {IRAS versus POTENT Density Fields on Large Scales: Biasing Parameter and
  Omega}, \apj, 495, 516

\bibitem[{{Silberstein}(1924)}]{Silberstein1924}
{Silberstein}, L. 1924, {The curvature of de Sitter's space-time derived from
  globular clusters}, \mnras, 84, 363

\bibitem[{{Slipher}(1913)}]{Slipher1913}
{Slipher}, V.~M. 1913, {The radial velocity of the Andromeda Nebula}, Lowell
  Observatory Bulletin, 2, 56

\bibitem[{{Slipher}(1914)}]{Slipher1914}
--- 1914, {The detection of nebular rotation}, Lowell Observatory Bulletin, 2,
  66

\bibitem[{{Slipher}(1915)}]{Slipher1915}
--- 1915, {Spectrographic Observations of Nebulae}, Popular Astronomy, 23, 21

\bibitem[{{Slipher}(1917)}]{Slipher1917}
--- 1917, {Nebulae}, Proceedings of the American Philosophical Society, 56, 403

\bibitem[{{Sullivan}(1916)}]{Sullivan1916}
{Sullivan}, R. 1916, {Celestial motions in the line of sight}, Popular
  Astronomy, 24, 109

\bibitem[{{Tonry} et~al.(2003){Tonry}, {Schmidt}, {Barris}, {Candia},
  {Challis}, {Clocchiatti}, {Coil}, {Filippenko}, {Garnavich}, {Hogan},
  {Holland}, {Jha}, {Kirshner}, {Krisciunas}, {Leibundgut}, {Li}, {Matheson},
  {Phillips}, {Riess}, {Schommer}, {Smith}, {Sollerman}, {Spyromilio},
  {Stubbs}, \& {Suntzeff}}]{Tonry2003}
{Tonry}, J.~L., {Schmidt}, B.~P., {Barris}, B., {Candia}, P., {Challis}, P.,
  {Clocchiatti}, A., {Coil}, A.~L., {Filippenko}, A.~V., {Garnavich}, P.,
  {Hogan}, C., {Holland}, S.~T., {Jha}, S., {Kirshner}, R.~P., {Krisciunas},
  K., {Leibundgut}, B., {Li}, W., {Matheson}, T., {Phillips}, M.~M., {Riess},
  A.~G., {Schommer}, R., {Smith}, R.~C., {Sollerman}, J., {Spyromilio}, J.,
  {Stubbs}, C.~W., \& {Suntzeff}, N.~B. 2003, {Cosmological Results from High-z
  Supernovae}, \apj, 594, 1

\bibitem[{{Truman}(1916)}]{Truman1916}
{Truman}, O.~H. 1916, {The motions of the spiral nebulae}, Popular Astronomy,
  24, 111

\bibitem[{{Watkins} et~al.(2009){Watkins}, {Feldman}, \&
  {Hudson}}]{Watkins2009}
{Watkins}, R., {Feldman}, H.~A., \& {Hudson}, M.~J. 2009, {Consistently large
  cosmic flows on scales of 100$h^{-1}$Mpc: a challenge for the standard
  {$\Lambda$}CDM cosmology}, \mnras, 392, 743

\bibitem[{Weyl(1923)}]{Weyl1923}
Weyl, H. 1923, Raum. Zeit. Materie: Vorlesungen {\"u}ber allgemeine
  Relativit{\"a}tstheorie (J. Springer)

\bibitem[{{Wirtz}(1916)}]{Wirtz1916}
{Wirtz}, C. 1916, {Die Trift der Nebelflecke}, Astronomische Nachrichten, 203,
  293

\bibitem[{{Wirtz}(1917)}]{Wirtz1917}
--- 1917, {{\"U}ber die Eigenbewegungen der Nebelflecke}, Astronomische
  Nachrichten, 204, 23

\bibitem[{{Wirtz}(1924)}]{Wirtz1924}
--- 1924, {De Sitters Kosmologie und die Radialbewegungen der Spiralnebel},
  Astronomische Nachrichten, 222, 21

\bibitem[{{Young} \& {Harper}(1916)}]{Young1916}
{Young}, R.~K., \& {Harper}, W.~E. 1916, {The Solar Motion as Determined from
  the Radial Velocities of Spiral Nebul{\ae}}, \jrasc, 10, 134

\end{thebibliography}

\end{document}